\documentclass[fleqn,usenatbib]{mnras}

\usepackage[T1]{fontenc}

\DeclareRobustCommand{\VAN}[3]{#2}
\let\VANthebibliography\thebibliography
\def\thebibliography{\DeclareRobustCommand{\VAN}[3]{##3}\VANthebibliography}

\usepackage{graphicx}	
\usepackage{amsmath}	
\usepackage{amssymb}	
\usepackage{newtxtext,newtxmath}

\title[GJ~367: limits on binarity and giant exoplanets]{High-contrast, high-angular resolution view of the GJ~367 exoplanet system}
\author[W. Brandner, P. Calissendorff, N. Frankel, and F. Cantalloube]{
Wolfgang Brandner,$^{1}$\thanks{E-mail: brandner@mpia.de}
Per Calissendorff,$^{2}$
Neige Frankel,$^{3,1}$
and Faustine Cantalloube$^{4}$
\\
$^{1}$Max-Planck-Institut f\"ur Astronomie, K\"onigstuhl 17, 69117 Heidelberg, Germany\\
$^{2}$Department of Astronomy, University of Michigan, Ann Arbor, MI 4810, USA\\
$^{3}$Canadian Institute for Theoretical Astrophysics, University of Toronto, 60
St. George Street, Toronto, ON M5S 3H8, Canada\\
$^{4}$Aix Marseille Univ, CNRS, CNES, LAM, Marseille, France
}

\date{Accepted 2022 April 4. Received 2022 March 30; in original form 2022 February 3}

\pubyear{2022}

\begin{document}
\label{firstpage}
\pagerange{\pageref{firstpage}--\pageref{lastpage}}
\maketitle

\begin{abstract}
We search for additional companions in the GJ~367 exoplanet system, and aim at better constraining its age and evolutionary status.
We analyse high contrast direct imaging observations obtained with HST/NICMOS, VLT/NACO, and VLT/SPHERE. We investigate and critically discuss conflicting age indicators based on theoretical isochrones and models for Galactic dynamics. 
A comparison of GAIA EDR3 parallax and photometric measurements with theoretical isochrones suggest a young age $\le$60\,Myr for GJ~367. The star's Galactic kinematics exclude membership to any nearby young moving group or stellar stream. Its highly eccentric Galactic orbit, however, is atypical for a young star. Age estimates considering Galactic dynamical evolution are most consistent with an age of 1 to 8\,Gyr. We find no evidence for a significant mid-infrared excess in the WISE bands, suggesting the absence of warm dust in the GJ~367 system. 
The direct imaging data provide significantly improved detection limits compared to previous studies. At 530\,mas (5\,au) separation, the SPHERE data achieve a 5 sigma contrast of $2.6 \times 10^{-6}$. The data exclude the presence of a stellar companion at projected separations $\ge$0.4\,au. At projected separations $\ge$5\,au we can exclude substellar companions with a mass $\ge$ 1.5 M$_{\rm Jup}$ for an age of 50\,Myr, and $\ge$ 20 M$_{\rm Jup}$ for an age of 5\,Gyr.  
By applying the stellar parameters corresponding to the 50\,Myr isochrone, we derive a bulk density of $\rho_{\rm planet} = 6.2$\,g/cm$^3$ for GJ\,367~b, which is 25\% smaller than a previous estimate.
\end{abstract}





\section{Introduction}

GJ~367 is an early M-dwarf located at a distance of 9.4\,pc, with a close-to-solar metallicity of $[\rm{Fe/H}] \approx -0.01 \pm 0.12$. It is host to GJ~367b, a transiting exoplanet with a 7.7\,h orbital period, corresponding to a semi-major axis of $\approx$0.007\,au ($\approx$ 0.75 mas maximum projected separation for a circular orbit). With an estimated radius of 0.72 R$_{\rm Earth}$ and mass of 0.55{\raisebox{0.5ex}{\tiny$^{+0.03}_{-0.04}$}} M$_{\rm Earth}$, GJ~367b is currently the smallest and lowest mass exoplanet known within 10\,pc of the Sun. The implied density of 8.2 g/cm$^3$ classifies it as a rocky planet with an extended, iron dominated core. With a dayside equilibrium temperature of 1500 to 1750\,K, GJ~367b classifies as a lava planet \citep{Lam2021}.

The host star GJ~367 is showing no signs of strong variability or activity. Photometric monitoring covering 5\,yr, and comprising 11 distinct observing epochs, derived a V-band variability amplitude of 12\,mmag \citep{Hosey2015}. GJ~367's activity index of R$' _{HK} = -5.15 \pm 0.12$ and rotational period of P$_{rot}=53$\,day \citep{Astudillo2017} closely follow the relation typical for relatively quiet early M-dwarfs \citep{Suarez2015}. 
One peculiarity of GJ~367 is its high space motion with respect to the Sun. As recently as 130,000 yr ago, the Sun and GJ~367 had their closest encounter at a perihelion distance of $\approx$5.4\,pc \citep{BailerJones2018}. An analysis combining HIPPARCOS and GAIA EDR3 astrometry of GJ~367 finds a $3.8 \sigma$ significance for a proper motion anomaly, possibly indicating the presence of a substellar companion \citep{Kervella2022}.
The GAIA EDR3 astrometric excess noise, which is the noise to be added to the individual GAIA observations for achieving a reduced $\chi ^2$ of 1 in the astrometric fit of the single star model \citep{Lindegren2018}, has a significance of $\approx 17\sigma$ \citep{GAIA2016,GAIA2021}.

The proximity to the Sun, and the EDR3 astrometric anomaly make GJ~367 an interesting target for a  direct imaging search for companions at projected separations $\gtrapprox$ 1\,au. 

The outline of the paper is as follows: In section 2 we present the high-angular resolution data and their analysis. In section  3 we investigate various age indicators and the Galactic kinematics. In section 4 we discuss the findings, and identify four challenges towards a better understanding of the GJ~367 exoplanet system. 


\section{Observations and data reduction}

We identified three high angular and high contrast observations of GJ~367 in the Mikulski Archive for Space Telescopes and the archive of the European Southern Observatory (see Table \ref{ObservationLog}).

\begin{table*}
\caption{Basic parameters of high contrast observations of GJ~367}             
\label{ObservationLog}      
\centering                          
\begin{tabular}{l| c c c c }        
Telescope/instrument &HST/NICMOS& VLT/NACO  &\multicolumn{2}{c}{VLT/SPHERE} \\ 
Camera & NIC2 & S13 & IFS & IRDIS\\
Field of view [$'' \times ''$] &19.4 $\times$ 19.4 &13.6 $\times$ 13.6 &1.73 $\times$ 1.73  &11.0 $\times$ 12.5  \\
Image scale [mas/pixel] &75.65 & 13.25 &7.46&12.25 \\
Filter/wavelength range & F110W, F180M, F207M, F222M & F185N & YJ &  H2+H3 \\
Angular resolution [mas] &$\gtrapprox$95 &48&$\geq$25$^*$ &$\gtrapprox$41$^*$ \\
Total integration time [s] & 64, 64, 128, 128 & 280 &3136 & 2816 \\
Integration set-up & STEP64, STEP64, STEP128, STEP128 & 0.5s $\times$ 20 $\times$ 28 &32s $\times$ 1 $\times$ 98 & 8.0s $\times$ 2 $\times$ 176 \\
Telescope tracking & stare & auto-jitter & \multicolumn{2}{c}{pupil stabilized} \\
Programme ID & GO 7420 & 70.C-0738(A) & \multicolumn{2}{c}{0104.C-0336(A)} \\
Principal Investigator & David A.\ Golimowski & Jean-Luc Beuzit & \multicolumn{2}{c}{Anna Boehle} \\
Observing date & 1997-09-17 & 2003-03-17 & \multicolumn{2}{c}{2020-02-08} \\
Remarks & snapshot & low wind effect & \multicolumn{2}{c}{ADI + Lyot coronagraph} \\
\hline                                   
\end{tabular}
     \begin{quote}
        $^*$Inner working angle $\approx$150\,mas with apodized Lyot coronagraph.
      \end{quote}
\end{table*}

\begin{table*}
\caption{Companion candidates (cc) to GJ~367}             
\label{CompanionCandidates}      
\centering                          
\begin{tabular}{l| c r r r c c}        
ID & Epoch & sep [$''$]& PA [deg] & $\Delta$mag [mag] & status &GAIA EDR3 ID \\ \hline
cc1 & 1997-09-17 &8.03 &222.2 &11.46 (F110W) &non-common proper motion & 5412250575032741504 \\
cc2 & 1997-09-17 &8.45 &3.1 & 3.39 (F110W) &non-common proper motion & 5412250575040992256 \\
cc3 & 1997-09-17 &11.54 &18.9 &11.95 (F110W) &background star (no CH$_4$ absoprtion) &--- \\
cc4 & 1997-09-17 &8.72 &40.7 &10.07 (F110W) &non-common proper motion & 5412250575040992768 \\
cc5 & 1997-09-17 &9.05 &51.8 & 8.44 (F110W) &non-common proper motion & 5412250540683548160 \\
cc6 & 2020-02-08 &6.48 &39.8 & 11.13 (H2) &background M-dwarf (H2-H3 colour) &--- \\
\hline                                   
\end{tabular}
     \begin{quote}
        Separation ``sep'' and Position Angle ``PA'' are measured relative to GJ~367 in the epoch of the respective high-contrast observations.
      \end{quote}
\end{table*}

\subsection{HST/NICMOS}

\begin{figure*}
    \begin{center}
        \includegraphics[width=0.98\textwidth]{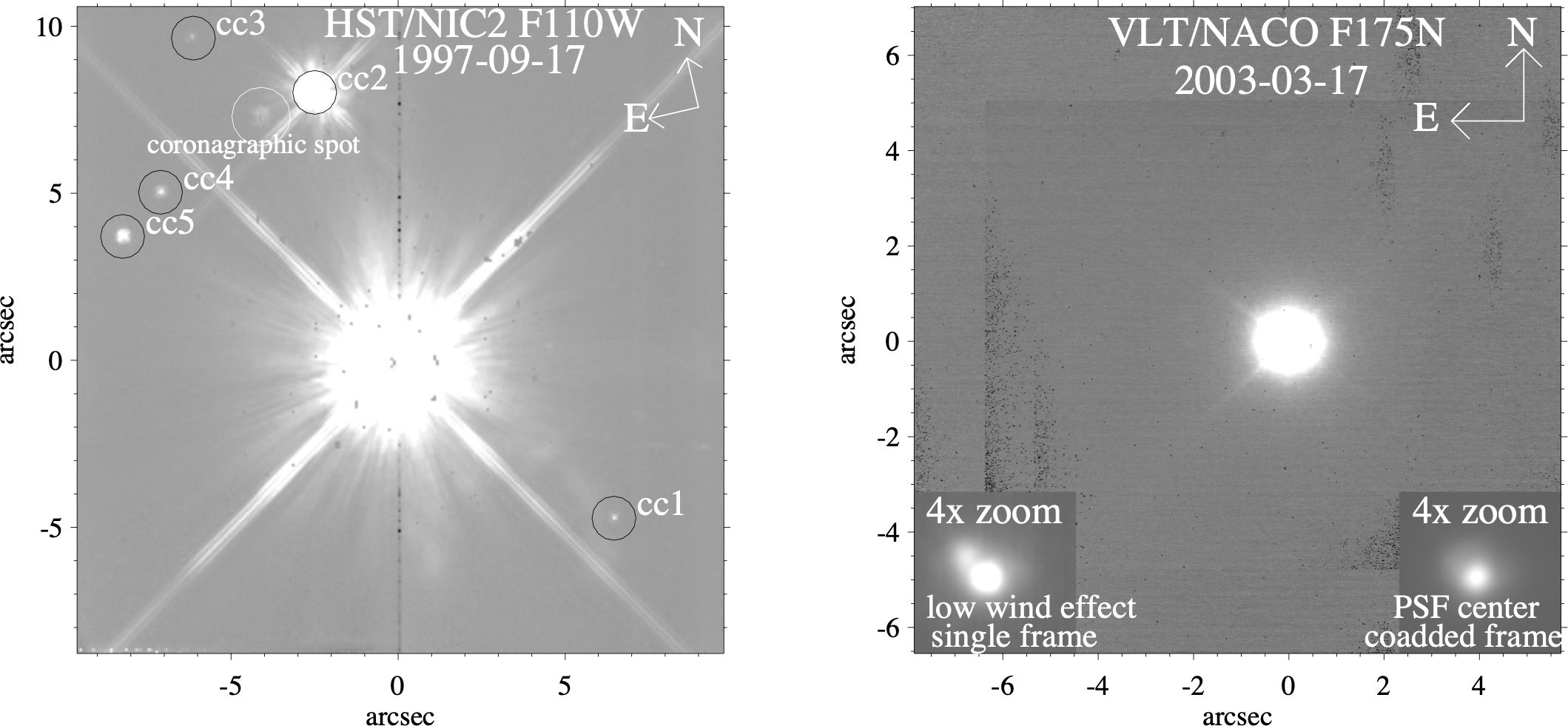}
    \end{center}
    \caption[]{Left: HST NIC2 image in F110W centered on GJ~367, with five companion candidates (cc) visible (black circles). The location of residuals originating in NIC2's coronagraphic mask is indicated (white circle). Right: VLT/NACO image in NB175 centered on GJ~367. The lower left insert shows a $4\times$ zoom-in on an individual 10s frame depicting a side-lope resulting from the low wind effect. The insert on the lower right shows a $4\times$ zoom-in of the coadded PSF.}
    \label{GJ367_NIC2_NACO}
\end{figure*}

GJ~367 was observed with HST/NICMOS \citep{Thompson1998} as part of a snapshot survey for companions to nearby stars (see Table \ref{ObservationLog} for observational details). The filter combination had been selected to provide a differentiator between background stars and substellar companions with CH$_4$ absorption bands \citep{Dieterich2012}. 

We detect 5 companion candidates (cc) to GJ~367 in all 4 bands of the NIC2 data set (see Figure \ref{GJ367_NIC2_NACO}, left, cc1 to cc5, and Table \ref{CompanionCandidates}). GAIA parallax measurements and upper limits identify four of these sources as background objects. The brightest source cc2 was already classified as a background dwarf star based on its F110W - F222M colour by \cite{Dieterich2012}. The 5th and faintest source cc3 has no counterpart in EDR3. The fact that it is detected in all 4 NIC2 filters suggests the absence of CH$_4$ absorption bands, thus it is most likely also an unrelated background source.

\cite{Dieterich2012} provide radial detection limits in F180M, which reach 3 to 5 $\sigma$ contrast ratios $\gtrapprox 2.4 \times 10^4$ for angular separations $\ge 3''$. The HST/NIC2 radial contrast limits for  separations $\le 2''$ ($\lessapprox$20\,au projected separation) are included in Figure \ref{contrast_h2}.  


\subsection{VLT/NACO}

GJ~367 was observed with VLT/NACO \citep{Lenzen2003,Rousset2003} as part of a programme to probe for stellar multiplicity among nearby low mass stars (see Table \ref{ObservationLog}). 
The observations were carried out during a period of very low ground level winds (<v$_{\rm wind}$> = 0.37$\pm$0.19 m/s). The resulting low-wind effect is known to create phase shifts in the wavefront in specific quadrants of the telescope pupil, resulting in one or multiple (transient) side-lopes of point sources \citep{Milli2018,Sauvage2016}. 
We reduced the data with the EsoRex NACO pipeline Version 4.4.10. Figure \ref{GJ367_NIC2_NACO} (right) shows the coadded image. The insert on the lower left shows as an example one of several individual frames affected by the low-wind effect. The insert in the lower right shows the coadded frame. The side-lope is much less pronounced, yet still visible as a diffuse ``smudge'' to the upper left of the PSF center.

Compared to the shortest wavelength HST/NIC2 observations, the VLT/NACO observations have a 2 times finer diffraction-limited resolution. The data confirm that GJ~367 is unresolved (i.e.\ not a stellar binary) down to a projected separation of $\approx$0.4\,au (Figure \ref{GJ367_NIC2_NACO}, right).

We also  carried out a visual search for companion candidates. No other sources were detected down to m$_{\rm NB175} \approx$ 13.5\,mag (i.e.\ about 7.5\,mag fainter than GJ~367). cc1, which is about 10.1 mag fainter than GJ~367 in GAIA BP and RP, and like GJ~367 has GAIA BP - RP $\approx$2.3\,mag, is below the detection limit of the VLT/NACO data set.


\subsection{VLT/SPHERE}


\begin{figure*}
    \begin{center}
        \includegraphics[width=0.98\textwidth]{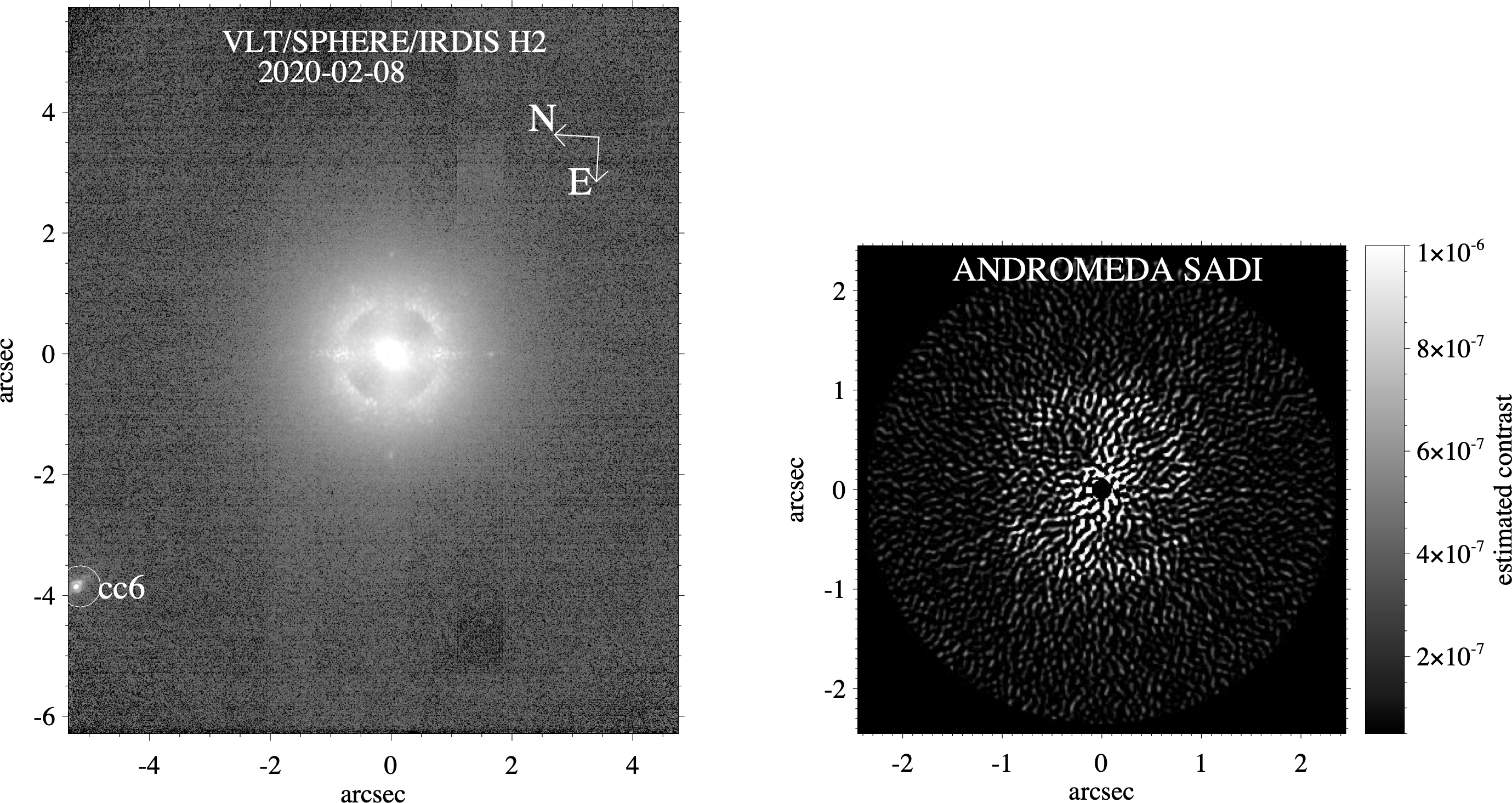}
    \end{center}
    \caption[]{Left: the companion candidate cc6 is detected at the very edge of the IRDIS field of view. Right: map of the estimated contrast for the IRDIS H2+H3 observations following SADI processing with ANDROMEDA, based on the assumption that a point source would be located at this pixel position.}
    \label{GJ367_IRDIS_SADI}
\end{figure*}

\begin{figure}
    \begin{center}
        \includegraphics[width=0.48\textwidth]{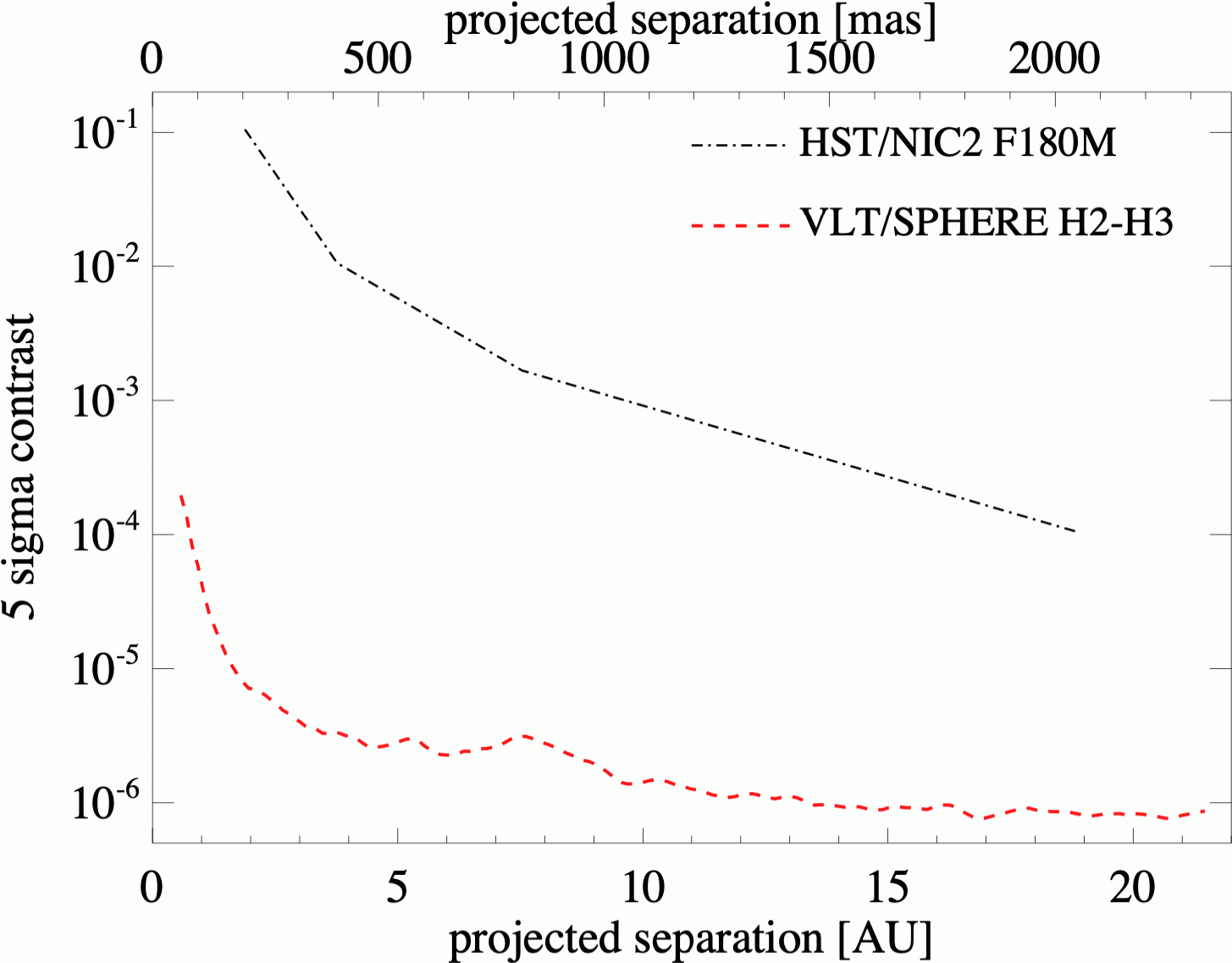}
    \end{center}
    \caption[]{Radial 5\,$\sigma$ contrast achieved with NIC2 in the F180M filter (dash-dotted line, \cite{Dieterich2012}), and SADI processed SPHERE/IRDIS in the H2/H3-filters (red dashed line) plotted over the projected separation.}
    \label{contrast_h2}
\end{figure}

GJ~367 was observed with VLT/SPHERE \citep{Beuzit2019} as part of a programme to probe for substellar companions to nearby stars (see Table \ref{ObservationLog}).
The pupil stabilized angular differential imaging \citep[ADI]{Marois2006} observing sequence was obtained with the apodized Lyot coronagraph \citep{Soummer2011, Carbillet2011, Martinez2009}.
Before the start and after the end of the coronagraphic observing sequence a single data set with the satellite spots turned on was recorded in order to facilitate the estimation of the star center position behind the coronagraph. During the ADI sequence, the satellite spots were not present.

We reduced the data using a combination of the EsoRex SPHERE pipeline version 0.42.0 and the vlt-sphere python package version 1.4.3 \citep{Vigan2020}\footnote{https://github.com/avigan/SPHERE}. For the astrometric calibration (i.e.\ image plate scale and orientation) we adopted the values by \cite{Maire2016}.

A point source cc6 is detected (Figure \ref{GJ367_IRDIS_SADI}, left, and Table \ref{CompanionCandidates}) in the IRDIS H2 ($\lambda_{\rm c} = 1589$\,nm) and H3 ($\lambda_{\rm c} = 1667$\,nm) bands. This source has no counterpart in EDR3. Compared to GJ~367, cc6's H2-H3 colours are  $\approx 0.1$\,mag redder. Thus most likely it is a background star of slightly later spectral type than GJ\,367. Due to the high proper motion of GJ~367, there is no overlap between the field of view of NIC2 in 1997 and IRDIS in 2020.

The resulting pre-reduced 4D (x,y,time,wavelength) IRDIS and IFS data sets were further processed with ANDROMEDA version 3.1 \citep{Cantalloube2015}. Figure \ref{GJ367_IRDIS_SADI} (right) shows the resulting residual after spectral ADI (SADI) processing of the IRDIS data. No close companion candidate was detected. Figure \ref{contrast_h2} shows the radial contrast curve (red dashed line) for our SADI analysis of the SPHERE/IRDIS dual band imaging data. In the central 700\,mas ($\approx$6.6\,au projected separation), IFS and IRDIS contrast curves after SADI processing are quite similar. 
VLT/SPHERE achieves a 5$\sigma$ contrast of $\approx 2.6 \times 10^{-6}$ at 5\,au (520 mas), and $\approx 1 \times 10^{-6}$ beyond 14\,au (1500 mas).
The HST/NIC2 contrast curve (black dotted line) according to \cite{Dieterich2012} is overplotted for reference. 


\section{Age and formation history of GJ~367}

\begin{table}
\caption{Compilation of (apparent) optical, near-, and mid-infrared photometry, distance modulus (DM), and absolute flux (i.e.\ at 10\,pc distance) of GJ~367}             
\label{Photometry}      
\centering                          
\begin{tabular}{l c c c}        
Band& [mag] &$\lambda$F$_\lambda$ [10$^{-12}$\,W/m$^2$] &source \\ \hline
G$_{\rm BP}$&10.3735$\pm$0.0008 & 1.346$\pm$0.002  & EDR3\\
G           &9.1587$\pm$0.0003 & 3.069$\pm$0.002 & EDR3\\
G$_{\rm RP}$&8.0582$\pm$0.0005  &  4.959$\pm$0.003  & EDR3\\
J &6.632$\pm$0.023 & 7.635$\pm$0.007    & 2MASS\\
H &6.045$\pm$0.044 &  6.26$\pm$0.13   & 2MASS\\
Ks &5.780$\pm$0.020&   4.00$\pm$0.17   & 2MASS\\
W1 & 5.550$\pm$0.164&  1.459$\pm$0.027   & AllWISE\\
W2 &5.364$\pm$0.061 & 0.71$\pm$0.12    & AllWISE\\
W3 &5.477$\pm$0.015 & 0.045$\pm$0.003    & AllWISE\\
W4 &5.381$\pm$0.038 & 0.0071$\pm$0.0001    & AllWISE\\ \hline
DM & $-0.13006${\raisebox{0.5ex}{\tiny$^{+0.00058}_{-0.00029}$}}   & --- & EDR3\\ \hline                        
\end{tabular}
     \begin{quote}
        References: EDR3 - \cite{GAIA2021}; 2MASS - \cite{Cutri2003}; AllWISE - \cite{Cutri2014}
      \end{quote}
\end{table}

The age determination presents the main challenge in converting the contrast ratio between a substellar companion and its host star into a mass estimate. Primary age indicators are the evolutionary state of the star, its activity and rotational period, and Galactic kinematics \citep{Soderblom2010}.

\subsection{Isochronal age and metallicity}

\begin{figure*}
    \begin{center}
        \includegraphics[width=0.49\textwidth]{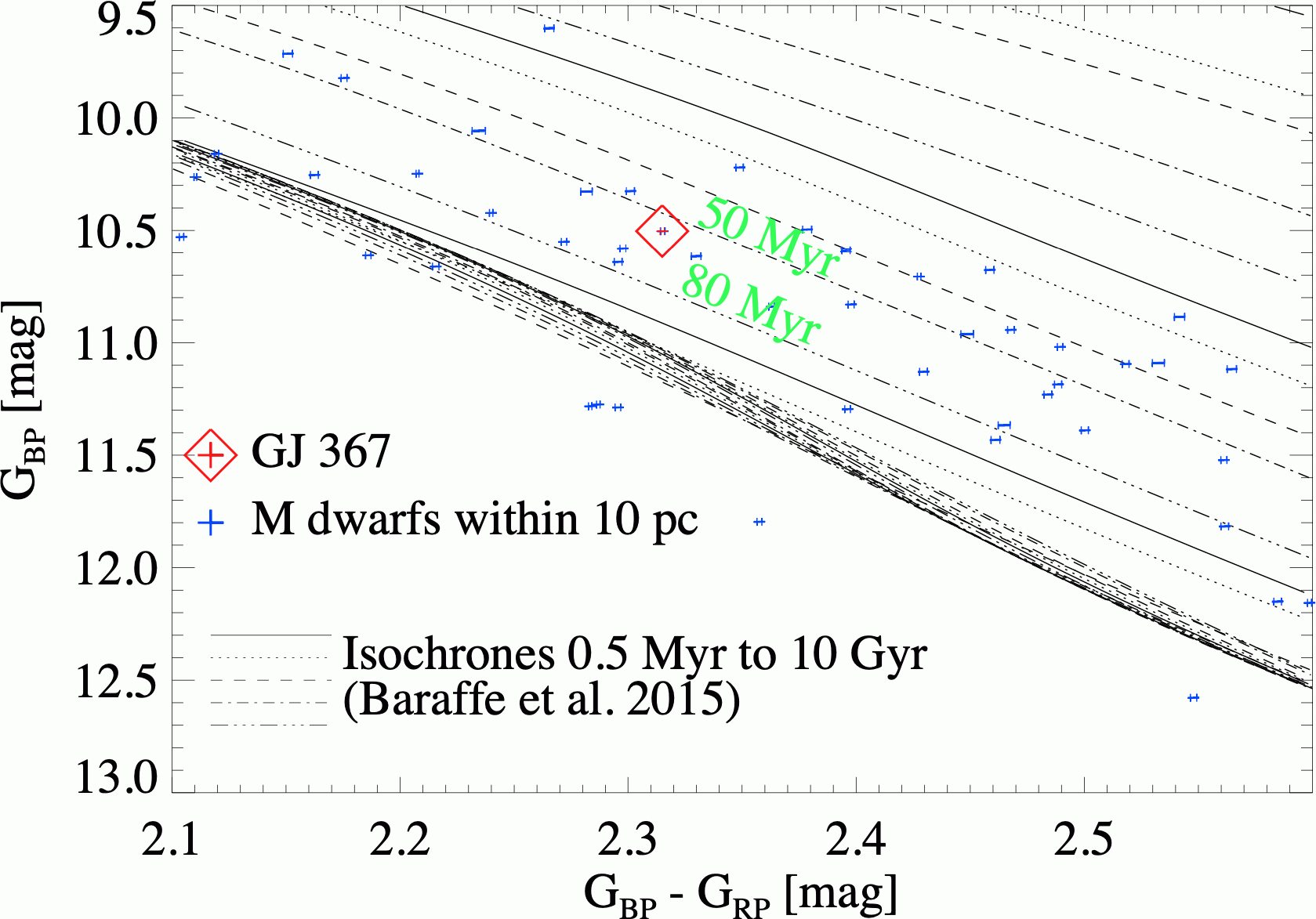}
        \includegraphics[width=0.49\textwidth]{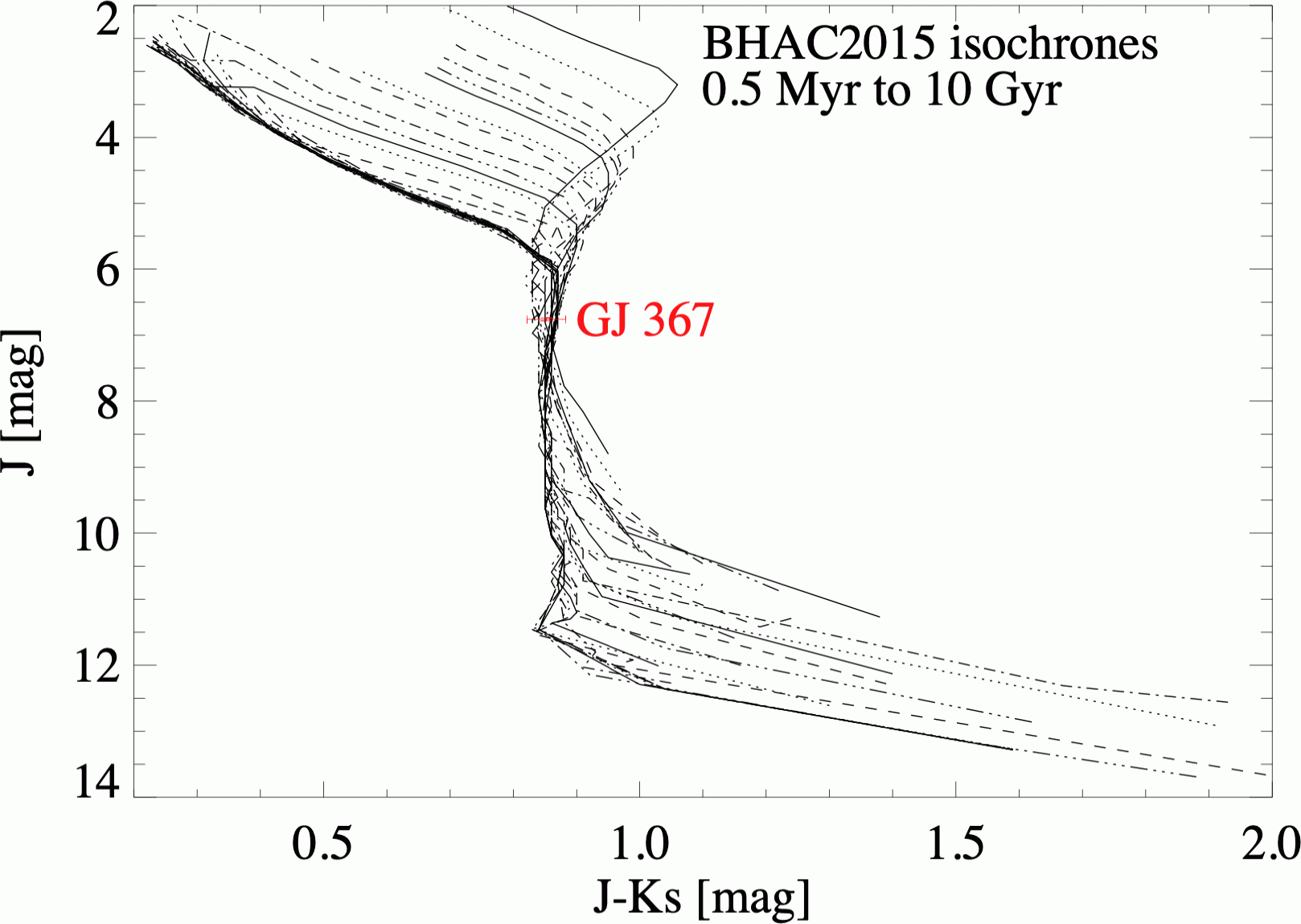}
    \end{center}
    \caption[]{Colour-magnitude diagrams for GJ~367 compared of theoretical models by \cite{Baraffe2015}. The plot on the left shows isochrones in the GAIA photometric systems, with GJ~367 (red symbol) being located below the 50\,Myr and above the 80\,Myr isochrone. The plot on the right illustrates that isochrones transformed to 2MASS bands do not provide any meaningful constraint on GJ~367's age. The 1\,$\sigma$ error bars in both plots are based on the uncertainties in photometry and parallax.}
    \label{GJ367_age_mag}
\end{figure*}

\begin{table}
\caption{Compilation of metallicity estimates of GJ~367}             
\label{gj367_metal}      
\centering                          
\begin{tabular}{c l}        
[Fe/H]& reference \\ \hline
$-0.07\pm0.09$& \cite{Kuznetsov2019}\\
$-0.06\pm0.09$& \cite{Maldonado2019}\\
$-0.01\pm0.12$& \cite{Lam2021}\\ \hline
\end{tabular}
\end{table}

The proximity of GJ~367 results in precise parallax measurements, and hence a precise determination of its absolute magnitudes. Table \ref{Photometry} presents a compilation of optical and infrared photometry of GJ~367. Its metallicity is close to solar \citep{Lam2021}, which makes theoretical isochrones computed for solar metallicity directly applicable (see Table \ref{gj367_metal} for a compilation of recent metallicity estimates). In Figure \ref{GJ367_age_mag}, left, we compare GJ~367's absolute GAIA BP magnitude and GAIA BP - RP colour with theoretical isochrones by \cite{Baraffe2015}. Under the assumption that GJ~367 is a single star, the isochrones suggest an age of around 55 to 60\,Myr. Isochrones in the 2MASS photometric system, on the other hand, do not provide an age discriminant. This is in part due to the larger photometric uncertainties of the 2MASS measurements, and in part due to the fact that the isochrones in the 2MASS system are degenerate\footnote{As pointed out by \cite{Allard1995}, for late-type dwarfs of solar metallicty, molecular opacities lock in place the peak of the flux at $\approx$1.1\,$\mu$m over a broad range of effective temperatures. Thus the flux in the visual provides a better diagnostics on effective temperature and evolutionary state.} for objects of GJ~367's near infrared colours and magnitudes (Figure \ref{GJ367_age_mag}, right). Thus we consider the isochronal age estimate of $8.0${\raisebox{0.5ex}{\tiny$^{+3.8}_{-4.6}$}}\,Gyr by \cite{Lam2021}, which is based on the GAIA DR2 parallax and the 2MASS photometry, of limited informative value.

A comparison of GJ~367's location in a GAIA colour-magnitude diagram with the other $\approx$150 M dwarfs within 10\,pc of the Sun \citep[using the sample defined by][]{Stelzer2013} reveals that its colour and magnitude is rather typical (i.e.\ average) for an early M-dwarf (Figure \ref{GJ367_age_mag}, left). The dispersion in absolute magnitudes above the (solar metallicity) main sequence for nearby late K- and early M-type stars can in part be explained by unresolved binary stars \citep{Zari2018}, and in part by a spread in metallicity above solar. 

By being about 0.5\,mag brighter in the optical and near infrared than the 1 to 10 Gyr solar-metallicity main-sequence, GJ~367's location in the colour magnitude diagram coincides with the binary sequence. Our direct imaging data exclude a stellar companion at projected separations $\ge$0.4\,au in the 2003 data set, and at projected separations $\ge$1\,au in the data sets obtained in 1997 and 2020. We also found no evidence in the literature for GJ~367 having a closer-in stellar companion (i.e.\ no hint for a spectroscopic, astrometric, or eclipsing binary).

In order to investigate the effect of the uncertainty in the metallicty of GJ~367 (Table \ref{gj367_metal}) on the isochronal age estimate, we use the Mesa Isochrones and Stellar Tracks \citep[MIST,][]{Dotter2016} in the GAIA EDR3 photometric system and for $[\rm{Fe/H}] = (-0.25, 0.00, +0.25$). 
To account for the long rotational period of GJ~367, we selected the isochrones for v/v$_{\rm crit}$=0.0. For $[\rm{Fe/H}] = 0.0 \pm 0.25$, we derive an isochronal age of $28${\raisebox{0.5ex}{\tiny$^{+35}_{-17}$}}\,Myr for GJ\,367. In general, sub-solar metallicity isochrones suggest a younger age, and super-solar metallicity isochrones yield an older age. 

According to the MIST models, a 1 to 10\,Gyr old early M-dwarf with a metallicity of [Fe/H] $\approx +0.45$ would fit GJ\,367's optical and near infrared luminosity. Such a high metallicity, though, deviates by at least 4\,$\sigma$ from GJ\,367's close-to-solar metallicity estimates. 

\subsection{Mid-infrared emission}

\cite{Bentley2019} investigated GAIA, 2MASS, and WISE photometry of $\approx$100,000 nearby M-dwarfs. With G-J = 2.53\,mag, and G-K = 3.38\,mag, GJ\,367's GAIA and 2MASS colours are typical for M1 to M2 dwarfs. Compared to early M-dwarfs in the $\ge$500\,Myr old field star sample, though, GJ\,367's reveals a 2 to 3\,$\sigma$ infrared excess in the WISE bands (K-W2 = 0.42\,mag, W1-W2=0.19\,mag, see Table \ref{Photometry}).
We also compare the spectral energy distribution of GJ\,367 with models for a 0.4\,M$_\odot$ star by \cite{Baraffe2015}. According to the models, the 0.4\,M$_\odot$ star has T$_{\rm eff}$ = 3512\,K at an age of 50\,Myr, and T$_{\rm eff}$ = 3521\,K at an age of 5\,Gyr, which is a very good match to the T$_{\rm eff}$ = 3522$\pm$70\,K reported by \cite{Lam2021}. While the 50\,Myr model provides a good fit to the photometry of GJ~367, the 5\,Gyr model predicts only about 65\% to 70\% of its observed optical, near- and mid-infrared flux. 

\subsection{Gyrochronology and age}

Gyrochronology relies on the assumption that young stars at the end of their contraction phase arrive on the main-sequence with a high-angular momentum, and subsequently are able to lose angular momentum by stellar winds. Main-sequence stars with longer rotational periods should thus be on average older than main sequence stars with shorter rotational period. This then facilitates age-dating of a main-sequence star from its rotational period \citep{Mamajek2012}. 

Based on photometric monitoring data, \cite{Lam2021} deduced a rotational period of $48\pm 2$\,d for GJ~367, which is consistent with the rotational period of P$_{rot}=53$\,d according to \cite{Astudillo2017}. \cite{Lam2021} derived a gyrochronological age of $4.0\pm 1.1$\,Gyr.

For early M-dwarfs, a study of single and multiple stars in the Hyades finds a significant dearth of rapid rotators around single stars with masses $\ge$0.3\,M$_\odot$ \citep{Douglas2016}. Single Hyades members from the sample defined by \cite{Kopytova2016} with masses of $\approx$0.4\,M$_\odot$ have a typical rotational period of $\approx$20\,days, which is about 3 times longer than predicted by models of the angular momentum evolution of low mass stars \citep{Reiners2012}. According to \cite{Douglas2016} this suggests that for single stars with undisturbed circumstellar disks, magnetic braking might be a highly efficient mechanism to shed angular momentum. As a consequence the rotational period of early M-dwarfs might have less informative power as an age indicator.

\subsection{Galactic kinematics and moving groups membership}

\begin{figure*}
    \begin{center}
        \includegraphics[width=0.95\textwidth]{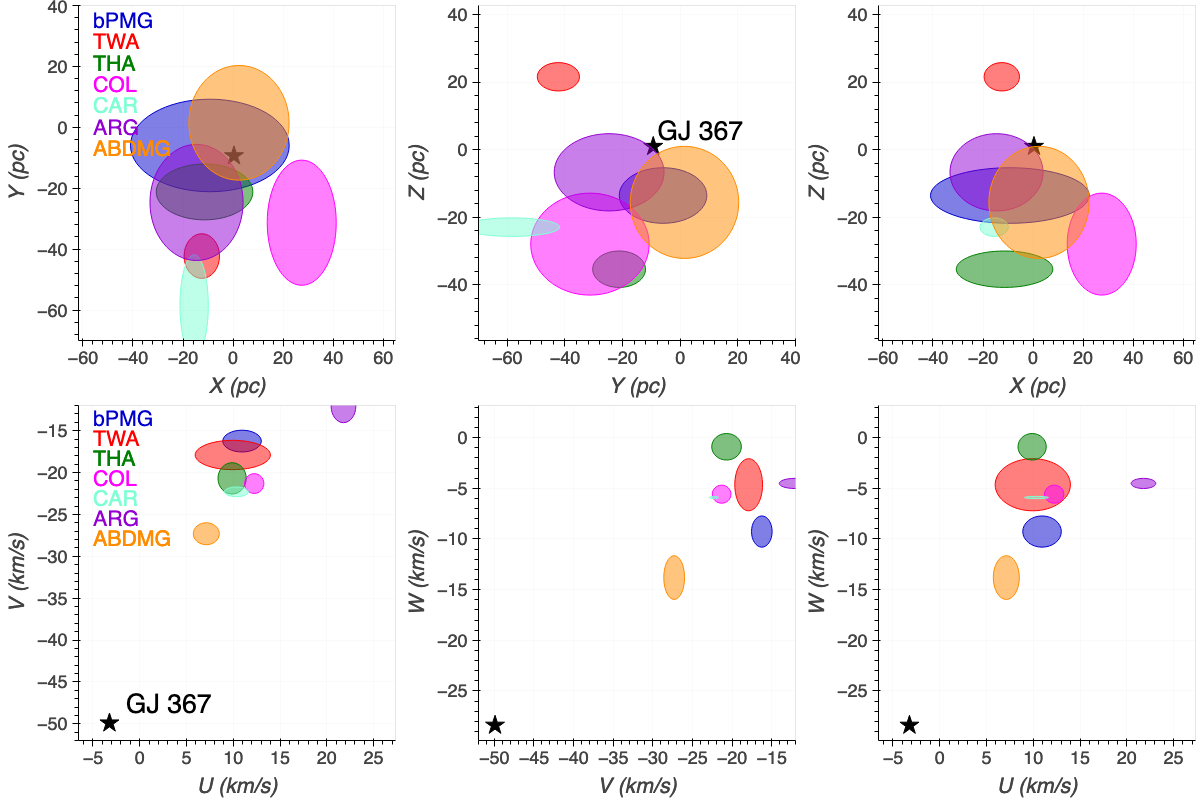}
    \end{center}
    \caption[]{XYZ coordinates and UVW velocities of GJ~367 compared to Young Moving Groups in a left-handed Galactic coordinate system (graphics adapted from \cite{Rodriguez2016})}
    \label{GJ367_YMG}
\end{figure*}

\begin{table}
\caption{GJ~367's Galactic coordinates and velocities in a left-handed coordinate system (not corrected for LSR)}             
\label{xyzuvw}      
\centering                          
\begin{tabular}{c|c}        
XYZ [pc]& $0.3667$, $-9.3621$, $0.9289$ \\
UVW [km/s]& $-3.27$, $-50.87$, $-28.30$\\ \hline
\end{tabular}
     \begin{quote}
        Note: Based on positions, parallax and proper motions from GAIA EDR3, radial velocity according to \cite{Trifonov2020}, and the transformation matrix defined in \cite{HIPPARCOS1997}. 
      \end{quote}
\end{table}

We base our investigation of GJ~367's Galactic kinematics on GAIA EDR3 positions, parallax and proper motions, and on the weighted mean radial velocity of RV = $47.9216\pm$0.0001\,km/s from the public HARPS data base corrected for systematics \citep{Trifonov2020}. The transformation from the GAIA EDR3 observables in the International Celestial Reference System (ICRS) to XYZ positions and UVW velocities (see Table \ref{xyzuvw}) in a left-handed Galactic reference frame (U positive for motions towards the Galactic anti-center) is based on the transformation matrix defined in \cite{HIPPARCOS1997}. The resulting values are U = $-3.3$ km/s, V = $-50.9$ km/s, and W = $-28.3$ km/s. This is in very good agreement with \cite{Lam2021} (U$_{\rm LSR}$, V$_{\rm LSR}$, W$_{\rm LSR}$  = (-11.73 $\pm$ 0.01, -36.5 $\pm$ 0.4, -21.93 $\pm$ 0.04)), when assuming that they based their correction to the local standard of rest (LSR) on \cite{Coskunoglu2011}, i.e.\ U$_\odot$, V$_\odot$, W$_\odot$ = (-8.50, 13.38, 6.49) km/s in a left-handed coordinate system (thus the minus sign in U$_\odot$).\footnote{We note that \cite{Lam2021}'s choice of LSR correction results in a small inconsistency when they assess the likelihood for GJ~367's membership to the different Galactic populations according to \cite{Bensby2003}. The latter report U$_{\rm LSR}$, V$_{\rm LSR}$, W$_{\rm LSR}$ for thin and thick disk, and halo stars in a right-handed coordinate system, and based on a solar motion with respect to the LSR of U$_\odot$, V$_\odot$, W$_\odot$ = (10.00, 5.25, 7.17) km/s according to \cite{Dehnen1998}. As \cite{Bensby2003} assume Gaussian probability distributions functions for the populations depending on the square of the UVW velocities, the choice of right- or left handed coordinate system does not affect the likelihood estimate, though the difference in particular in V$_\odot$ results in $\Delta$V$_{\rm LRS} \approx$8\,km/s.} 

Figure \ref{GJ367_YMG} illustrates that GJ~367's kinematics differ significantly from those of nearby young moving groups \citep{Malo2013}, and hence excludes membership to any of these groups. GJ~367's UVW velocities are also distinct from those of the Hercules stream \citep[see, e.~g.][]{Chen2021}. This finding is supported by the LACEwING \citep{Riedel2017} and BANYAN \citep{Gagne2018} analysis tools for assessing membership to nearby young moving groups. Thus GJ~367 is not a member of any nearby moving group or stream.

\subsection{Age constraints from Galactic dynamics}

\begin{figure}
    \begin{center}
        \includegraphics[width=0.48\textwidth]{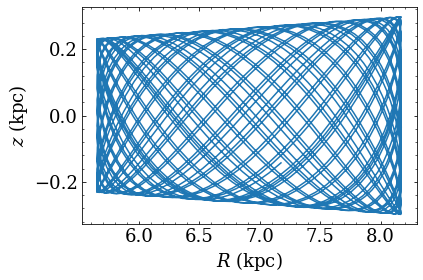}
    \end{center}
    \caption[]{Vertical-radial projection of multiple Galactic orbits of GJ~367. At the solar circle, GJ~367 is close to its apocenter, with the pericenter corresponding to R$\approx 5.65$\,kpc. GJ~367 oscillates with an amplitude of $\pm$240 to $\pm$280\,pc vertical to the Galactic plane.}
    \label{GJ367_orbits}
\end{figure}

\begin{figure}
    \begin{center}
        \includegraphics[width=0.48\textwidth]{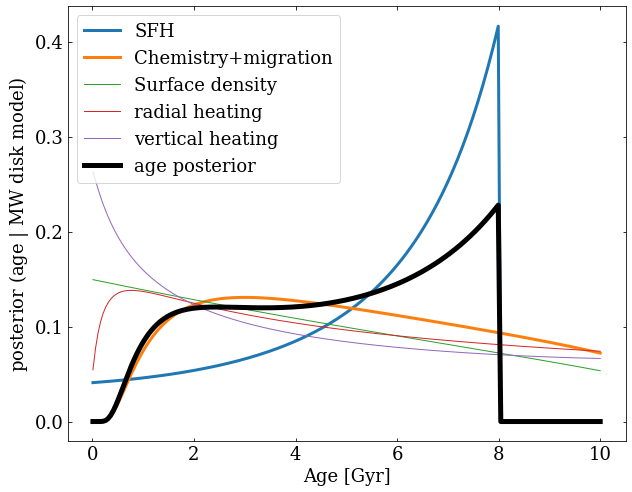}
    \end{center}
    \caption[]{Posterior age of GJ~367 based on [Fe/H], Galactic angular momentum, radial action, and height above the Galactic plane (thick black). Contributions from individual model components entering in the posterior are re-normalized and over-plotted according to the legend: star formation history and radial surface density, chemical information, orbital heating.}
    \label{GJ367_ageposterior}
\end{figure}

The birth conditions and subsequent evolutionary history of stellar populations determine their present-day Galactic kinematics and metallicity. \cite{Frankel2018,Frankel2020} constructed a global model for the distribution of stars' [Fe/H], ages, positions and velocities, and fitted it to APOGEE DR12 and DR14 data. Here, we use this model to produce an age posterior for the star, given its [Fe/H] and kinematics.

Noteworthy are GJ~367's V and W components, i.e., the velocity component parallel to the Galactic plane and perpendicular with respect to the Galactic center, and the velocity component vertical to the Galactic plane. Relative to the Sun, GJ~367's V component is $\approx$50 km/s less, and thus considerably slower than the circular velocity at the position of the Sun. This indicates a Galactic orbit with a significant eccentricity. 

We used galpy v1.7.1 \citep{Bovy2015} to integrate GJ~367's orbit with the MWPotential2014 Galactic potential. Figure \ref{GJ367_orbits} shows the results of the integration in the radial-vertical plane. We selected the following plausible input parameters: R = 8.12 kpc \citep{GRAVITY2018}, v$_{\rm T}$ = -14.5 km/s, v$_{\rm R}$ = 198 km/s, z =  22 pc, v$_{\rm z}$ = -21 km/s. 
According to the orbit integration, the pericenter and apocenter of GJ~367's orbit are at $\approx$5.65 and $\approx$8.15\,kpc, corresponding to an eccentricity e$\approx$0.3.\footnote{We aim at a qualitative analysis. Quantitative estimates will vary according to the precise choice of the LSR correction.}

The following analysis (Figure \ref{GJ367_ageposterior}) is along the lines of the investigation carried out in order to age-date the HD 19467 system \citep{Maire2020}. It is based on the latest version of the Galactic models as described in \cite{Frankel2020}.
The analysis derives an age posterior considering 1) the Galactic star formation history (SFH), 2) the chemical and migration model (i.e.\ the stellar metallicity and position), 3) the surface density of stars, 4) the radial heating, which is reflected in the eccentricity of the stellar orbit, and 5) the vertical heating, i.e.\ the average height of the star above the Galactic plane.

The most influential contributions are i) the star formation history of the low-alpha disk, which describes that there are more old than young stars, ii) the chemical+migration model, which --- given this star's metallicity and position --- provides a range of possible birth places and birth times. If there would be no radial migration in the Milky Way, this would be the most constraining parameter. For GJ~367, it {\b shows a broad and flat peak between 1.5 and 4 Gyr}. Thus the star formation history still dominates the posterior.

Another aspects is the (currently) low vertical height of the star. As a consequence, the vertical secular evolution model favours a young age. The eccentricity, on the other hand, is quite large. Thus the in-plane secular evolution favours an older age.

Overall, the disk model produces a very broad posterior for the age of the star, with probable values between 1 and 8 Gyr. The chemo-dynamical aspects of the model tend to favour a young age, whereas the global star formation history aspect draws the posterior to and older age. Utilizing additional chemical information in the future (e.g.\ including alpha element abundances) may help constraining the age further.

\begin{table}
\caption{Age estimators for GJ~367}             
\label{gj367_age}      
\centering                          
\begin{tabular}{l c c c}        
method& age [Myr] & reference \\ \hline
Isochrones: BHAC$^a$&57{\raisebox{0.5ex}{\tiny$^{+3}_{-2}$}}  & this paper\\
Isochrones: MESA$^b$&$28${\raisebox{0.5ex}{\tiny$^{+35}_{-17}$}}  & this paper\\
Gyrochronology&$3980\pm1110$ &\cite{Lam2021}\\
Galactic Dynamics$^c$&$5500${\raisebox{0.5ex}{\tiny$^{+2500}_{-4500}$}}&this paper\\ \hline
\end{tabular}
     \begin{quote}
        $^a$ assuming solar metallicity for GJ~367; $^b$ assuming $-0.25\le$[FeH]$\le +0.25$ for GJ~367; $^c$ assuming that GJ~367's space motion is the result of many low momentum exchange encounters, and not dominated by a single, high momentum exchange n-body interaction. Note: given the spread of age estimates, we decided to choose 50\,Myr and 5\,Gyr as exemplary ages for the conversion of contrast limits into mass limits.
      \end{quote}
\end{table}


\section{Discussion}

\begin{figure}
    \begin{center}
        \includegraphics[width=0.48\textwidth]{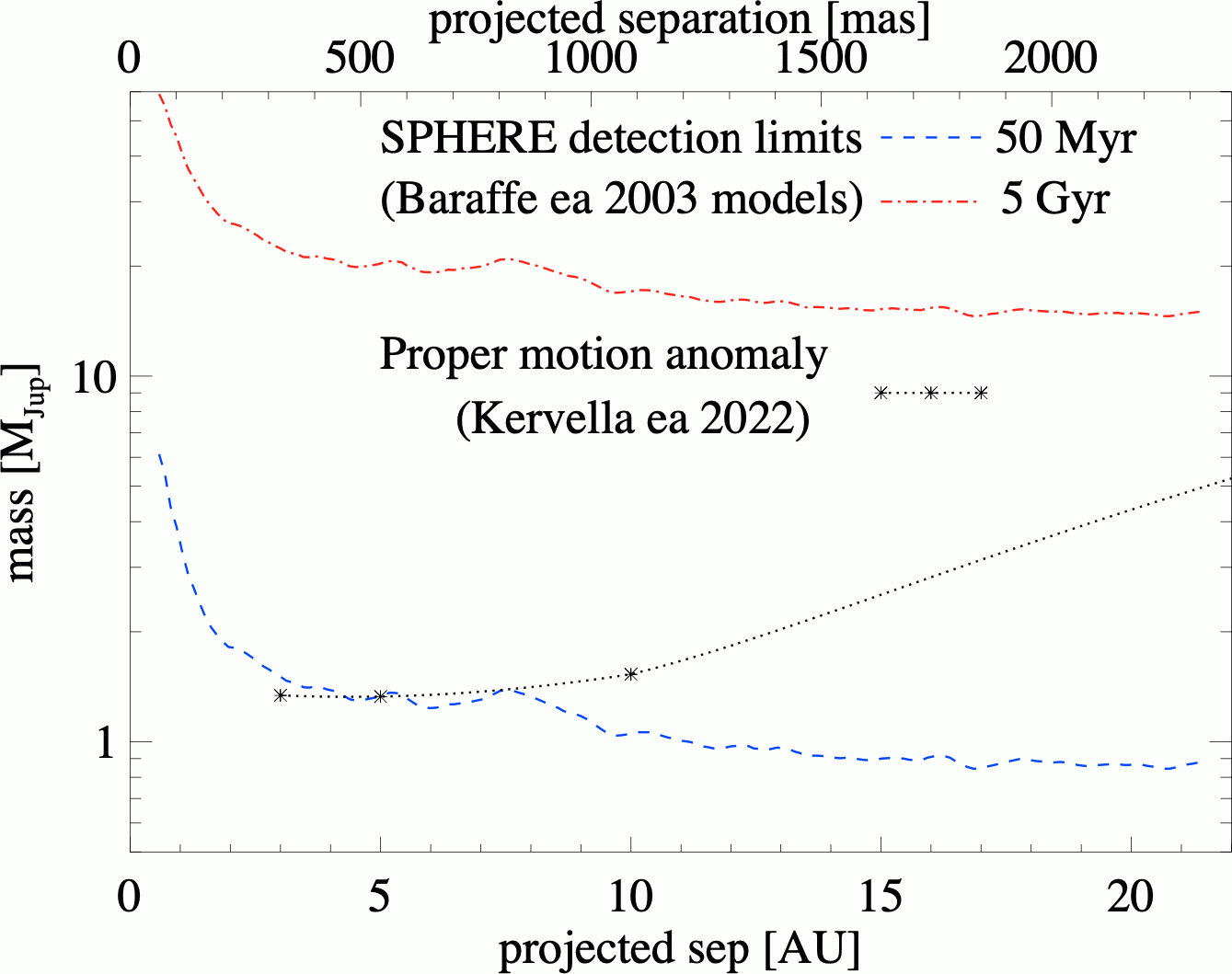}
    \end{center}
    \caption[]{Radial mass detection limits for SPHERE, transformed from 5\,$\sigma$ contrast limits using the models by \cite{Baraffe03} for ages of 50\,Myr (blue dashed line) and 5\,Gyr (red dashed-dotted line). The asterisk symbols and the dotted line mark the companion mass to orbital radius relation explaining the proper motion anomaly of GJ~367 according to \cite{Kervella2022}.}
    \label{mass_radius}
\end{figure}

The VLT/SPHERE observations provide significantly improved detection limits compared to previous direct imaging observations of the GJ~367 system. The corresponding companion mass detection limits according to models by \cite{Baraffe03} for ages of 50\,Myr and 5\,Gyr are shown in Figure \ref{mass_radius}. 
The presence of a massive brown dwarf companion is excluded even for relatively old system ages. For a young system age, the observations would have been sufficiently sensitive to detect a 1.5\,M$_{\rm Jup}$ planet at a projected separation of 5\,au. This could help to narrow down the parameter range of possible companion masses and separations calculated by \cite{Kervella2022}. According to their study, the HIPPARCOS - GAIA EDR3 proper motion anomaly would require a companion with a mass $\ge$1.5\,M$_{\rm Jup}$ orbiting with a semimajor axis $\ge$10\,au, or a companion with a mass $\approx$1.3\,M$_{\rm Jup}$ orbiting with a semimajor axis $\lessapprox$5\,au (see Figure \ref{mass_radius}).

The exoplanet host GJ~367 provides contradicting age indicators (Table \ref{gj367_age}). Its intrinsic over-luminosity compared to 1 to 10\,Gyr old early M-dwarfs of solar metallicity suggests a maximum age of 60\,Myr.

Direct imaging observations rule out blending with a background star, or a stellar companion in a wider ($\gtrapprox 1$\,au) orbit. Radial velocity monitoring also rules out a closer-in stellar companion as an explanation for the over-luminosity. Metallicity measurements rule out the high metallicity of [Fe/H] $\approx +0.45$ required according to the MIST models for a 1 to 10\,Gyr old early M-dwarf to match GJ\,367's GAIA EDR3 parallax and photometric measurements. 

GJ~367's Galactic kinematics, and in particular the eccentricity of its Galactic orbit are well explained by Galactic dynamical evolution over a time period of 1 to 8\,Gyr. Such an age is in agreement with the gyrochronological age, but in conflict with the isochronal age. Future studies of the [$\alpha$/Fe] abundance of GJ~367 might provide more stringent constraints on its membership to a particular Galactic stellar population.

An alternative explanation for the eccentricity of GJ~367's Galactic orbit would be a fairly recent high-energy gravitational encounter, which changed its Galactic orbital parameters. This encounter could have happened early in the youth of GJ~367, and would point to its formation in a non-hierarchical multiple system, or a relatively dense cluster environment. While we cannot distinguish between these two scenarios, there is a high probability that either event would have significantly disturbed the orbits of any potential companions to GJ~367. This would be one explanation for GJ~367 lacking companions in wide orbits. In general, the formation environment  leaves an imprint on the architecture of planetary systems \citep{Winter2020}.

In the following we identify four challenges. Their solution might provide a better understanding of the age and evolutionary history and status of the GJ~367 exoplanet system.

\begin{itemize}
\item Challenge 1: can GJ~367 be overluminous, and not be young?
\item Challenge 2: can magnetic braking explain the long rotational period of 50 days for an early M dwarf with an age $\leq$60 Myr?
\item Challenge 3: does the integration of GJ~367's Galactic orbit back in time by 60 Myr hint at a possible birth place in a non-hierarchical multiple system, or a relatively dense stellar cluster?
\item Challenge 4: does the EDR3 astrometric excess noise hint at the presence of another substellar objects in the GJ~367 system? Or is the astrometric excess noise related to the limitation of the single star linear astrometric fit in EDR3, and GJ~367's proximity and peculiar space motion with respect to the Sun?\footnote{After correcting for the projection of linear proper motion onto the celestial sphere, \cite{Brandt2021} finds a good agreement between GJ~367's proper motion derived from GAIA EDR3 and HIPPARCOS  positions, and the GAIA EDR3 proper motion.}

\end{itemize}

Significantly deeper direct imaging detection limits might have to wait for the next generation of telescopes and instruments. Contrast ratios of a few times 10$^{-9}$ as envisioned for the Roman Space Telescope and its Coronagraph Instrument \citep{Trauger2016,Kasdin2020,Carrion2021} could push detection limits to sub-Saturn mass planets.\footnote{GJ\,367 might be on the faint side of stars accessible by the Roman Coronagraph Instrument.} 

An application of the stellar parameters according to the models by \cite{Baraffe2015} (BHAC) for an age of 50\,Myr changes the dependent properties of GJ\,367~b. The best fitting stellar model has a mass of 0.4\,M$_\odot$, T$_{\rm eff}$=3512\,K, and r$_{\rm star}$ = 0.479\,R$_\odot$. While the effective temperature agrees very well with the value determined by \cite{Lam2021}, the models describe a star with a 5\% larger radius, and a 10\% lower mass. An increase in the planetary radius 
from 0.72 to 0.75\,R$_{\rm Earth}$, and a decrease in the planetary mass from 0.55 to 0.48\,M$_{\rm Earth}$
would result in a $\approx$25\% decrease in the bulk density to $\rho_{\rm planet} = 6.2$g/cm$^3$ for GJ\,367~b.

Irrespective of its age, GJ\,367\,b is a benchmark rocky planet in the solar neighbourhood. While the semimajor axis of $\approx$0.007\,au of GJ\,367\,b is too close to its host star for direct imaging observations, phase curve and primary and secondary transit measurements should provide interesting insights into its day- and nightside properties. At a young age, the entire surface of GJ\,367\,b might still be covered by lava, while at an older age the nightside might have cooled below the melting point. Thus the day- to nightside temperature contrast of 
GJ\,367\,b could serves as another age indicator for the system. At a young age, GJ\,367\,b could also serve as a template to Earth's Hadean period \citep{Bonati2019}. 


\section*{Acknowledgements}

We thank H.\ Zinnecker for helpful comments on a draft of the paper.

Based on observations collected at the European Organisation for Astronomical Research in the Southern Hemisphere under ESO programmes 70.C-0738(A) and 0104.C-0336(A).

Based on observations made with the NASA/ESA Hubble Space Telescope (GO 7420), obtained from the data archive at the Space Telescope Science Institute. STScI is operated by the Association of Universities for Research in Astronomy, Inc. under NASA contract NAS 5-26555.

This work has made use of data from the European Space Agency (ESA) mission
{\it Gaia} (\url{https://www.cosmos.esa.int/gaia}), processed by the {\it Gaia}
Data Processing and Analysis Consortium (DPAC,
\url{https://www.cosmos.esa.int/web/gaia/dpac/consortium}). Funding for the DPAC
has been provided by national institutions, in particular the institutions
participating in the {\it Gaia} Multilateral Agreement.

This publication makes use of data products from the Two Micron All Sky Survey, which is a joint project of the University of Massachusetts and the Infrared Processing and Analysis Center/California Institute of Technology, funded by the National Aeronautics and Space Administration and the National Science Foundation.

This publication makes use of data products from the Wide-field Infrared Survey Explorer, which is a joint project of the University of California, Los Angeles, and the Jet Propulsion Laboratory/California Institute of Technology, funded by the National Aeronautics and Space Administration.

NF was supported by the Natural Sciences and Engineering Research
Council of Canada (NSERC), [funding reference number CITA 490888-16] through the CITA postdoctoral fellowship and acknowledges partial support from a Arts \& Sciences Postdoctoral Fellowship at the University of Toronto.

\section*{Data availability}

The data underlying this article are available online from the Barbara A. Mikulski Archive for Space Telescopes and the archive of the European Southern Observatory. The final data products are made available by contacting  the corresponding author.



\bibliographystyle{mnras}
\bibliography{lit} 


\bsp	
\label{lastpage}
\end{document}